\documentclass[12pt]{iopart}

\usepackage[latin1]{inputenc}
\usepackage{graphicx}
\begin{document}

\title[Surface critical behaviour of the interacting self-avoiding trail]{Surface critical behaviour of the Interacting Self-Avoiding Trail on the square lattice}

\author{D P  Foster}

\address{Laboratoire de Physique Th\'eorique et Mod\'elisation
(CNRS UMR 8089), Universit\'e de Cergy-Pontoise, 2 ave A. Chauvin
95302 Cergy-Pontoise cedex, France}

\begin{abstract}
The surface critical behaviour of the interacting self-avoiding trail is examined using transfer matrix methods coupled with finite-size scaling. Particular attention is paid to the critical exponents at the ordinary and special points along the collapse transition line. The phase diagram is also presented.
\end{abstract}

\maketitle

\section{Introduction}

Lattice self-avoiding walks have been used as models for real, linear polymers in solution for over three decades\cite{vanderbook}. The quality of the solvent may be introduced by the inclusion of short-ranged interactions in the model; typically an attractive energy is included for non-consecutive nearest-neighbour occupied lattice sites. This model is the standard Interacting Self-Avoiding Walk model (ISAW) or $\Theta$-point model\cite{flory, degennes75}. The model has been  shown to accurately predict the critical behaviour of a wide range of real linear polymers in solution, not only in the high-temperature phase, but also at the collapse transition, which occurs as  the temperature is lowered, at the $\Theta$ temperature.  The model is successful because it captures the strong entropic repulsion between different portions of the polymer chain (the self-avoidance), as well as the effect of the difference of affinity between monomer-monomer contacts and monomer-solvent contacts (attractive interaction). 

Whilst the relevant physical dimension in polymer physics would usually be $d=3$, the ISAW model has been much studied in two dimensions. This is partly motivated by the realisation that $d=3$ is the upper critical dimension of the collapse transition, and that the model in two dimensions provides an interesting playground.  In this paper we shall concentrate on the two-dimensional square lattice.

One could ask whether the ISAW is a special model, or whether other models which include the same basic ingredients would have the same critical behaviour.
 Two related models were introduced to examine this question: the O(n=0) model introduced by Nienhuis, which we will refer to in this paper as the Vertex-Interacting Self-Avoiding Walk  model 
 (VISAW)\cite{blotenienuis}, and the Interacting Self-Avoiding Trails model (ISAT)\cite{massih75}.  In both of these models the self-avoidance constraint is relaxed in that lattice sites may be visited more than once, but the lattice bonds may only be visited once. In the VISAW model the walk is not allowed to cross, but in the ISAT model the walk is allowed to cross. 

Whilst the VISAW and ISAT models have the same critical behaviour as the ISAW in the high-temperature phase\cite{blotenienuis,guim97}, the situation is different at the collapse transition. For the VISAW model, a mapping to an integrable 19 vertex model allows the exact calculation of the correlation exponent $\nu=12/23$ and the exponent $\gamma=53/46$\cite{wbn}, as compared to $\nu=4/7$ and $\gamma=8/7$ for the ISAW at the collapse transition\cite{ds}. The situation for the ISAT model is far less clear. Early studies found a number of contradictory results~\cite{lyklema,merovitchlim,OP95,OP07}. Some authors claimed that the results were compatible with the ISAT model and the ISAW model being in the same Universality class at the collapse transition\cite{merovitchlim}, whilst others found results which were incompatible\cite{OP95,OP07}. Recently this model was re-examined using transfer matrices\cite{F09}, and 
the conclusion arrived at was that the transition was similar to the collapse transition in the VISAW model (they have the same correlation length exponent $\nu$ but different $\gamma$~exponents). The collapse transition for both the VISAW and ISAT models is clearly not in the usual $\Theta$-point transition; the collapse transition in these models has an extra transition line leaving it separating two distinct dense phases. 

In our previous study of the ISAT model\cite{F09}, we gave results for the bulk-critical exponents and showed that these results were different from the results of Owczarek and Prellberg\cite{OP95}. We argued that this was due to a breakdown of the usual identification of the $\nu$ exponent with the geometrical exponent derived from the radius of gyration. In the current paper, we extend our study by introducing attractive interactions with a surface, and examine the surface critical behaviour of the model. In their paper\cite{OP95}, Owczarek and Prellberg also presented some results for the surface critical behaviour. Some of their results corresponded to explicit calculations using Monte Carlo, and others were derived by plausibility arguments. Our results concord with their explicit results, but disagree with the others. The surface critical exponents we find are consistent with the bulk results found in~\cite{F09}.

\section{Model and Transfer Matrix Calculation}

\begin{figure}
\begin{center}
\caption{A self-avoiding trail model showing the vertex crossings and 
the vertex collisions, both weighted with a Boltzmann factor $\tau$. Surface contacts are weighted $\omega_s$ and a fugacity $K$ is introduced per walk step. The trail is shown on a strip of width $L=5$. }\label{model}

\

\includegraphics[width=10cm,clip]{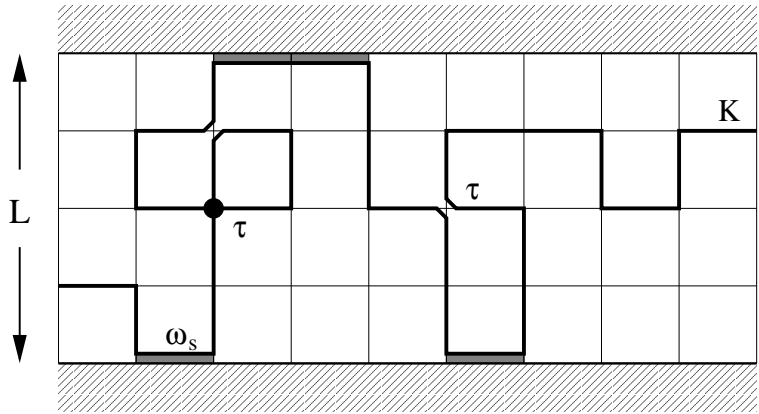}
\end{center}
\end{figure}

The ISAT model studied here is defined as follows: consider all random walks on the square lattice which do not visit any lattice bond more than once. Doubly visited sites may correspond to either crossings or ``collisions"; both are assigned an attractive energy $-\varepsilon$. The walk is allowed to touch, but not cross, a surface defined as a horizontal line on the lattice. Each step along the surface is assigned an attractive energy $-\varepsilon_S$. For the transfer matrix calculation that follows it is convenient to consider a strip of width $L$ with an attractive surface both sides of the strip. This is not expected to change the behaviour in the thermodynamic limit $L\to \infty$; the bulk critical behaviour should not depend on the boundary conditions and when calculating the surface critical behaviour, a walk adsorbed to one surface needs an infinite excursion in order to ``see" the other surface. Additionally, the finite-size scaling results which link the eigenvalues to of the transfer matrix to the scaling dimensions $x_\sigma^s$ and $x_\varepsilon^s$ (see \eref{sig-dim} and \eref{eng-dim}) rely on the conformal mapping of the half plane (one adsorbing surface) onto a strip with two adsorbing surfaces\cite{cardy}. A typical configuration for the ISAT is shown in \Fref{model}.

The partition function for the model is

\begin{equation}\label{part}
{\cal Z}=\sum_{\rm walks} K^N \omega_s^{N_S} \tau^{N_I},
\end{equation}
where $K$ is the fugacity, $\omega_s=\exp(\beta\varepsilon_S)$ and $\tau=\exp(\beta\varepsilon)$.
$N$ is the length of the walk, $N_S$ is the number of steps on the surface, and $N_I$ is the number of doubly-visited sites.

The average length of the trail is controlled by the fugacity $K$ through
\begin{equation}\label{n}
\langle N\rangle=K\frac{\partial \ln{\cal Z}}{\partial K}.
\end{equation}
As $K$ increases from zero, $\langle N \rangle$ increases, diverging at some value $K=K^*(\omega_s,\tau)$. To start we consider what happens in the absence of the adsorbing boundary. For $\tau$  small enough, 
\begin{equation}
\langle N\rangle\sim (K^*(\omega_s,\tau)-K)^{-1},
\end{equation}
 whilst for large enough $\tau$ the divergence is discontinuous. Whilst $\langle N\rangle$ is finite, the density of occupied bonds on an infinite lattice is zero, whilst once $\langle N \rangle$ has diverged the density is in general finite. For small enough $\tau$ the density becomes non-zero continuously at $K=K^*$ and for large enough $\tau$ the density jumps at $K=K^*$. $K^*$ may then be understood as the location of a phase transition, critical for $\tau<\tau_{\rm coll}$ and first order for $\tau>\tau_{\rm coll}$. The problem of infinite walks on the lattice is equivalent to setting $K=K^*$ and varying $\tau$, then it may be seen that for $\tau<\tau_{\rm coll}$ the density is zero and is non-zero for $\tau>\tau_{\rm coll}$. It then follows that $\tau_{\rm coll} $ defines the collapse transition point.
 

Now let us consider the effect of the adsorbing boundary at constant $\tau$. For $\omega_s$ small, the entropic repulsion of the wall is strong enough for the walk to remain in the bulk. Once $\omega_s$ is large enough for the energy gain to overcome the entropic repulsion, the walk will visit the boundary a macroscopic number of times, and the walk adsorbs to the surface. These two behaviours are separated by $\omega_s=\omega_s^*$. 
For $\omega_s\leq \omega_s^*$ the behaviour of the walk is not influenced by the wall, and $K^*$ is independent of $\omega_s$. The transition $K=K^*$ if critical ($\tau\leq\tau_{\rm coll}$) corresponds to ordinary critical behaviour. However, for $\omega_s>\omega_s^*$, $K^*$ is a function of $\omega_s$, and the transition is referred to as a surface transition. The point $K=K^*$, $\omega_s=\omega_s^*$ is referred to as the special critical point (again $\tau\leq\tau_{\rm coll}$). 

As the critical value $K^*$ is approached, and in the absence of a surface, the partition function \eref{part} and the correlation length $\xi$ are expected to diverge, defining the standard exponents $\gamma$ and $\nu$:
\begin{eqnarray}
\xi\sim|K-K^*|^{-\nu}\\
{\cal Z}\sim|K-K^*|^{-\gamma}
\end{eqnarray}
The effect of the surface on the walk is to introduce an entropic repulsion, pushing the walk away from the surface. The number of allowed walks is reduced exponentially if the walk is constrained to remain near the surface, in particular if one or both ends of the walk are obliged to remain in contact with the surface. In this case the divergence of $\cal Z$ is modified, and two new exponents are introduced, $\gamma_1$ and $\gamma_{11}$. Defining ${\cal Z}_1$ and  ${\cal Z}_{11}$ as the partition functions for a walk with one end, and both ends, attached to the surface respectively, then:
\begin{eqnarray}
{\cal Z}_1\sim|K-K^*|^{-\gamma_1}\\
{\cal Z}_{11}\sim|K-K^*|^{-\gamma_{11}}
\end{eqnarray}
Whilst the bulk exponents, such as $\nu$ and $\gamma$, are the same at an ordinary critical point and at the special critical point, the surface exponents $\gamma_1$ and $\gamma_{11}$ differ. 
The exponents $\nu$, $\gamma$, $\gamma_1$ and $\gamma_{11}$ are related by the Barber relation\cite{barber}:
\begin{equation}\label{barb}
\nu+\gamma=2\gamma_1-\gamma_{11}.
\end{equation}

This partition function may be calculated exactly on a strip of length $L_x\to\infty$ and of finite width $L$ by defining a transfer matrix ${\cal T}$. If periodic boundary conditions are assumed in the $x$-direction, the partition function for the strip is given by:
\begin{equation}
{\cal Z}_L=\lim_{L_x\to\infty}\Tr\left({\cal T}^{L_x}\right).
\end{equation}
The free energy per lattice site, the density, surface density and correlation length for the infinite strip may be calculated from the eigenvalues of the transfer matrix:
\begin{eqnarray}
f(K,\omega_s,\tau)=\frac{1}{L}\ln\left(\lambda_0\right),\\
\rho(K,\omega_s,\tau)= \frac{K}{L\lambda_0}\frac{\partial \lambda_0}{\partial K},\\
\rho_S(K,\omega_s,\tau)= \frac{\omega_s}{\lambda_0}\frac{\partial \lambda_0}{\partial \omega_s},\\\label{xi}
\xi(K,\omega_s,\tau)=\left(\ln\left|\frac{\lambda_0}{\lambda_1}\right|\right)^{-1},
\end{eqnarray}
where $\lambda_0$ and $\lambda_1$ are the largest and second largest (in modulus) eigenvalues.

Our first task is to find estimates of $K^*(\omega_s,\tau)$. 
For this, two distinct methods are used, which we now describe. 
\begin{enumerate}
\item
For $K\leq K^*$ the largest eigenvalue of the transfer matrix $\cal T$ corresponds to the empty lattice; the length of the walk is finite, but the lattice strip  is infinitely long, and so the probability of finding a non-empty column is zero. The largest eigenvalue is then $\lambda_0=1$. The divergence of the walk length is identified with the value $K=K^*_L$ for which $\xi\to\infty$. This occurs when 
$\lambda_1=\lambda_0=1$\cite{derh}. 
\item  An estimate for the critical point where the length of the walk diverges may be found using phenomenological renormalisation group for a pair of lattice widths\cite{mpn76}, $L$ and $L^\prime$. The estimated value of $K^*$ is given by the solution of the equation:
\begin{equation}\label{nrg}
\frac{\xi_L}{L}=\frac{\xi_{L^\prime}}{L^\prime}
\end{equation}
\end{enumerate}
Both these methods give finite-size estimates $K^*_L(\omega_s,\tau)$ which should converge to the same value in the limit $L\to\infty$.

The critical dimensions of the surface magnetic and energy fields may be calculated from the first few eigenvalues of the transfer matrix: 
\begin{eqnarray}\label{sig-dim}
x^s_\sigma&=&\frac{L\ln\left|\frac{\lambda_0}{\lambda_1}\right|}{\pi},\\\label{eng-dim}
x^s_\varepsilon&=&\frac{L\ln\left|\frac{\lambda_0}{\lambda_2}\right|}{\pi},
\end{eqnarray} 
with $\lambda_2$ the eigenvalue with the third largest absolute value.

The surface scaling dimensions $x^s_\sigma$ and $x^s_\varepsilon$ may be related to the surface correlation length exponent $\nu_s$ and the exponent $\eta_\parallel$, controlling the decay of the correlation function along the surface, through standard relations
\begin{eqnarray}\label{nuref}
\nu_s&=&\frac{1}{1-x^s_\varepsilon},\\\label{eta}
\eta_{\parallel}&=& 2x^s_\sigma.
\end{eqnarray}
The entropic exponent $\gamma_{11}$ is related to $\eta_{\parallel}$ through:
\begin{equation}\label{gam11eta}
\gamma_{11}=\nu(1-\eta_\parallel).
\end{equation}

For a more detailed discussion of the transfer matrix method, and in particular how to decompose the matrix, the reader is referred to the article of Blöte and Nienhuis~\cite{blotenienuis}.

\section{Results}

For clarity, we will present separately the results found setting $\lambda_1=1$ and those found using the phenomenological renormalisation group equation \eref{nrg}. 

The finite-size results obtained are, where possible, extrapolated on the one hand using the Burlisch and Stoer (BST) extrapolation procedure\cite{bst} and on the other hand fitting to a three point extrapolation scheme, fitting the following expression for quantity $X_L$:
\begin{equation}\label{3ext}
X_L=X_\infty+aL^{-b}.
\end{equation}
Calculating $X_\infty$, $a$ and $b$ require three lattice widths. The extrapolated values $X_\infty$ clearly will still depend weakly on $L$, and the procedure may be repeated, however weak parity effects can be seen in their values, impeding further reasonable extrapolation by this method.

\subsection{Results setting $\lambda_1=1$}

\begin{figure}
\begin{center}
\caption{$K_L^*(\omega_s,\tau)$ calculated setting $\lambda_1=1$ for $\tau=0$. This case corresponds to the adsorption of the standard SAW model.}\label{lamxt0}
\includegraphics[width=10cm,clip]{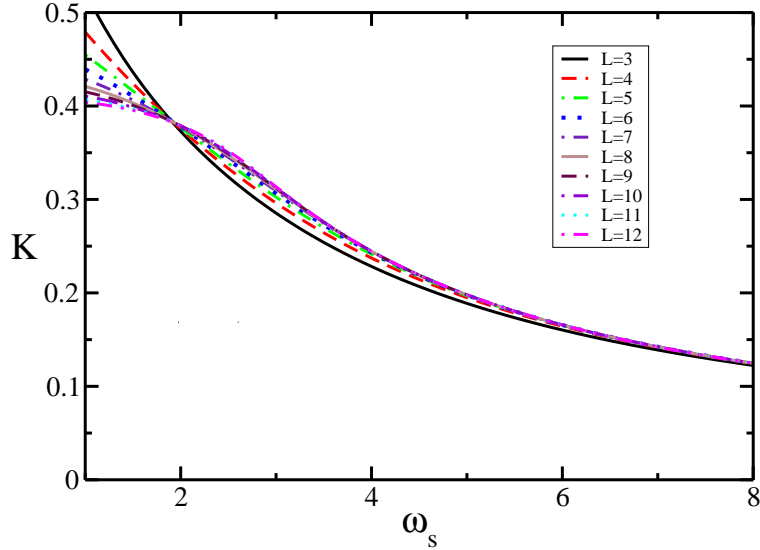}
\end{center}
\end{figure}

\begin{figure}
\caption{$K_L^*(\omega_s,\tau)$ calculated setting $\lambda_1=1$ for $\tau=3$. $\tau=3$ is expected to correspond to $\tau_{\rm coll}$, corresponding to the collapse transition line for $\omega_s<\omega_s^*$.}\label{lamxt3}
\begin{center}
\includegraphics[width=10cm,clip]{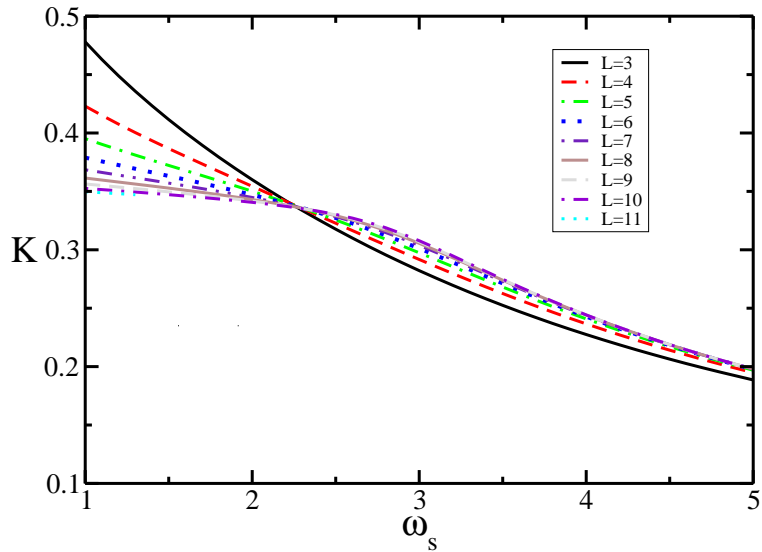}
\end{center}
\end{figure}

In \Fref{lamxt0}  we present the curves of $K_L^*(\omega_s,\tau)$ as a function of $\omega_s$ for $\tau=0$. The case where $\tau=0$ corresponds to the pure self-avoiding walk in the presence of an attractive surface, already studied using transfer matrices\cite{guimburk89,veal}.
The interacting self-avoiding trail maps onto the kinetic self-avoiding trail model when $\tau=3$. 
It was conjectured that the kinetic self-avoiding trail corresponded to the collapsing self-avoiding trail, leading to the identification
 $\tau_{\rm coll}=3$\cite{OP95}. This identification seems to be verified by the results presented in references\cite{OP95} and \cite{F09}.
The curves of $K_L^*(\omega_s,\tau)$ for $\tau=3$ are shown in \Fref{lamxt3}. The values calculated from the intersections of successive curves are shown in \Tref{Kcross} which places the special surface transition for $\tau=3$ at $K^*=0.3333\pm0.0001, \omega_s^*=2.427\pm0.002$. We calculated the finite-size estimates for the density $\rho_L(K^*,\omega_s^*)$. These estimates can be seen to have a finite limit of $\rho_\infty=0.35\pm0.02$. In \Tref{Kcross} we also give this reciprocal fractal dimension calculated at the special transition for $\tau=3$. Whilst the question of the density at the special point in the infinite lattice limit remains open, given the lattice widths used in our calculation, we can clearly see that we recover a geometrical exponent compatible with the results Owczarek and Prellberg\cite{OP95}.
This indicates that the  thermal exponent $\nu$ and geometrical exponent $1/d_f$ are likely to be 
different for this model at the collapse transition. 

\begin{figure}
\caption{Phase diagram in the $K=K_L^*(\omega_s,\tau)$ plane found setting $\lambda=1$ and identifying the adsorption transition with 
the the solutions of $K_L^*(\omega_s,\tau)=K_{L+1}^*(\omega_s,\tau)$.}\label{pdlam}
\begin{center}
\includegraphics[width=10cm,clip]{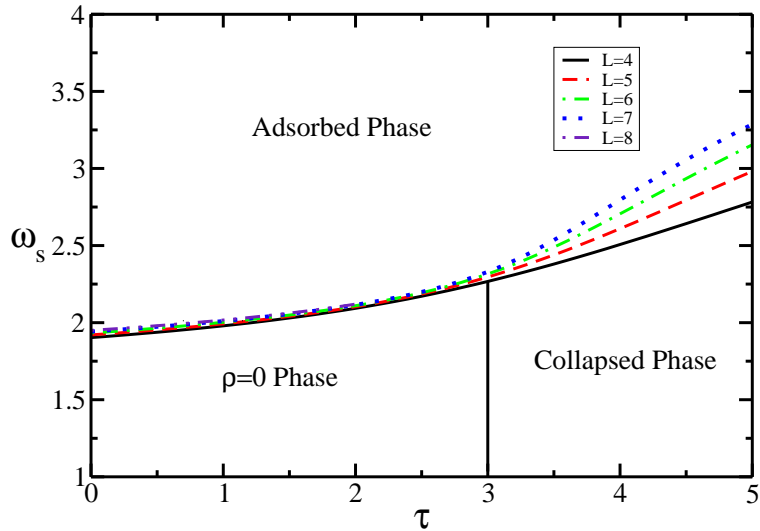}
\end{center}
\end{figure}

For both $\tau=0$ and $\tau=3$ the curves may clearly be seen to cross at a point, which defines the adsorption transition $\omega_s^*(\tau)$. By varying $\tau$ we may map out the adsorption line, and hence find the phase diagram in the $K=K^*(\omega_s,\tau)$ plane. This is shown in \fref{pdlam}. The vertical line, corresponding to the collapse transition, has been added by hand at $\tau_{\rm coll}=3$;
the boundary is not expected to influence the location of the collapse transition until the walk adsorbs to the boundary; the surface interaction below $\omega_s^*$ will have a finite perpendicular correlation length $\xi_\perp^s$ associated with it. In the thermodynamic limit, the bulk of the lattice will not be influenced surface, and bulk collapse should be unaffected. It is only at $\omega_s^*$, were $\xi^s_\perp\to\infty$, that the bulk will be influenced by the surface, and the walk will adsorb to the surface. This picture has been proven for the case of the adsorption of collapsing self-avoiding polygons in three dimensions\cite{Vrbova}.

\begin{table}
\caption{Location of the special point calculated by setting $\lambda_1=1$ and solving $K^*_L(\omega^*_s,\tau)=K^*_{L+1}(\omega_s^*,\tau)$ for $\tau=3$. The results have then been extrapolated using the BST extrapolation scheme. The reciprocal fractal dimension $1/d_f$ is given, calculated from the densities $\rho_L$ and $\rho_{L+1}$.}\label{Kcross}
\begin{indented}
\item[]\begin{tabular}{@{}lllll}
\br
$L$ & $\omega_s^*$ & $K^*$ & $\rho_{L+1}$ & $1/d_f$ \\
\mr
3 & 2.220342  &0.340394 & 0.501949 & 0.565358\\
4  &2.267276  &0.337428 & 0.474241 & 0.545825\\
5  &2.296043  &0.336003 &0.455622  & 0.534034\\
6  &2.315608  &0.335210 &0.442176 &  0.526218 \\
7  &2.329838  &0.334723 & 0.431958 & 0.520713 \\
8  &2.340687  &0.334403 & 0.423892 & 0.516664\\
9  &2.349254  &0.334182 & 0.417336 & 0.513588\\
10 & 2.356203  &0.334022 & 0.411881&0.511191\\
11 &2.361962  &0.333904  & 0.407254&0.509286\\
\mr
BST $\infty$ &  2.4267  &0.3333   &0.3539 & 0.498\\
\br
\multicolumn{4}{c}{Three point extrapolated results}\\
\mr
3 & 2.4227 & --- & --- & ---\\
4 & 2.4250 & 0.333267 & 0.327394 &  0.488281 \\
5 & 2.4262 & 0.333291 & 0.332348 &  0.490432 \\
6 & 2.4271 & 0.333304 & 0.334328 &   0.491926\\
7 & 2.4277 & 0.333312 &  0.334880 & 0.493045  \\
8 & 2.4281 & 0.333317 & 0.334636  & 0.493983 \\
9 & 2.4284 & 0.333321 & 0.333949 & 0.494649 \\
\br

\end{tabular}
\end{indented}
\end{table}

\begin{figure}
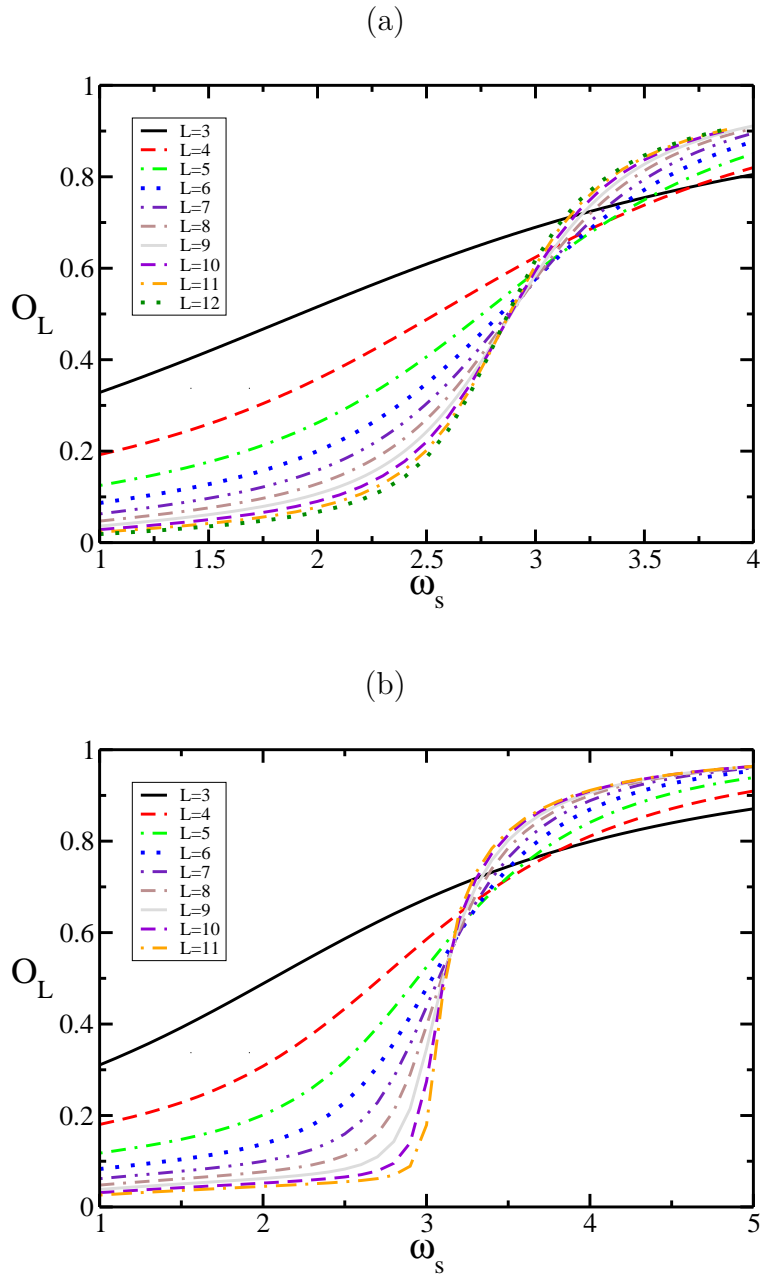

\caption{Plot of the order parameter ${\cal O}=\langle N_S\rangle/\langle N\rangle$ for (a) $\tau=3$ and (b) $\tau=4$, calculated setting $\lambda=1$.}\label{opfig}

\begin{center}
(a)

\ \\ 

\includegraphics[width=10cm,clip]{opt3}

\ \\ 

(b)

\ \\ 

\includegraphics[width=10cm,clip]{opt4}
\end{center}
\end{figure}
  
\begin{table}
\caption{Location of the special point calculated by setting $\lambda_1=1$ and solving $Y_{L,L+1}(\omega^*_s,\tau)=Y_{L+1,L+2}(\omega_s^*,\tau)={\cal Y}_L$ for $\tau=3$. The results have then been extrapolated using the BST extrapolation scheme. The precision of the estimates for ${\cal Y}_\infty$ is not sufficient for the BST extrapolation to be very accurate. The BST value is ${\cal Y}_\infty\approx 1.08$ which is a little low. Plotting the points, a reasonable estimate would be ${\cal Y}_\infty=1.12\pm 0.05$}\label{Ycross}
\begin{indented}
\item[]\begin{tabular}{@{}llll}
\br
$L$ &$\omega_s^*$ & $K^*$ & ${\cal Y}_L=(1-\phi_s)/\nu$ \\
\mr
3  &2.644273  &0.306760  &0.635164\\
4  &2.612544  &0.315671  &0.675046\\
5 &2.587753  &0.320837  &0.714237\\
6 &2.568090  &0.324099   &0.751030\\
7 &2.552274  &0.326280   &0.784704\\
8 &2.539271  &0.327805   &0.815357\\
9 &2.528203 & 0.328916   &0.843678 \\
10 &  2.519807 & 0.329713 & 0.866822\\
\mr
BST $\infty$  & 2.431 & 0.3339 &  see text and caption \\
\br
\end{tabular}
\end{indented}
\end{table}

 Having set $\tau=3$ using the conjecture from the kinetic SAT model, 
 there are different methods for estimating the location of the special surface transition, where collapse and adsorption occur simultaneously. 
The simplest is to use the crossings of $K^*_L$ to estimate the adsorption point $w_s^*$,
results for which are shown in \Tref{Kcross}. 
However, we may also use the scaling behaviour of the order parameter to set up a phenomenological renormalisation scheme. An order parameter for the adsorption transition may be defined as
\begin{equation}\label{op}
{\cal O}(K,\omega_s,\tau)=\frac{\langle N_S \rangle}{\langle N \rangle}.
\end{equation}
For a continuous adsorption transition, the order parameter $O$ vanishes for $\omega_s<\omega_s^*$
whilst for $\omega_s>\omega_s^*$ ${\cal O}$ is finite. 
At the transition point the order parameter is expected to scale as :
\begin{equation}
{\cal O}_L(\omega_s)=L^{-(1-\phi_s)/\nu}\tilde{O}\left[L^{y_s}(\omega_s-\omega_s^*)\right].
\end{equation}
Defining $Y_{L,L^\prime}$ by
\begin{equation}\label{ylog}
Y_{L,L^\prime}(\omega_s)=-\frac{\ln\left[{\cal O}_L(\omega_s)/{\cal O}_{L^{\prime}}(\omega_s)\right]}{\ln\left[L/L^\prime\right]},
\end{equation}
we may set up a phenomenological RG scheme for estimating the adsorption transition, as well as the exponent $\phi_s$. These are given by the solutions of the equation
\begin{equation}\label{yrg}
Y_{L,L+1}(\omega_s^*)=Y_{L+1,L+2}(\omega_s^*)={\cal Y}_L.
\end{equation}
where ${\cal Y}_L$ is the finite size estimate of the exponent ${\cal Y}_\infty=(1-\phi_s)/\nu$.
  
Plots of the order parameter are shown in \Fref{opfig} and results for ${\cal Y}_L$ are shown in \Tref{Ycross}. The values ${\cal Y}_L$ are far from their asymptotic values, and the extrapolation using  BST is not very conclusive. The best estimate from BST is ${\cal Y}_L=1.08$. Plotting this value along with the data points shows it to be a low estimate. A reasonable extrapolation of the points seems to be ${\cal Y}_L=1.12\pm 0.05$, although these values should be taken with care.

\subsection{Results from Phenomenological RG}

\begin{figure}
\begin{center}
\caption{$K_L^*(\omega_s,\tau)$ calculated using phenomenological RG for $\tau=3$.}\label{pdkwnrg}
\includegraphics[width=10cm,clip]{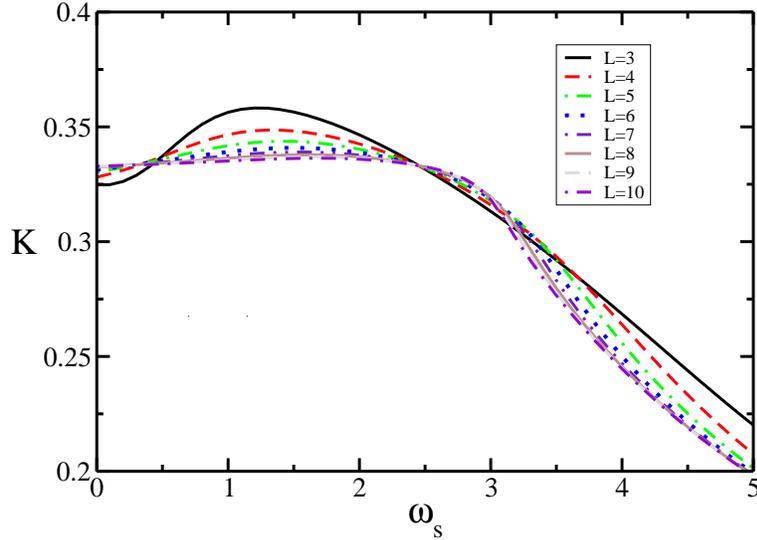}
\end{center}
\end{figure}

\begin{figure}
\caption{Exponent $\eta_\parallel$ from crossings of $\xi/L$ for $\tau=3$ and $L^\prime=L+1$}\label{eta11}
\begin{center}
\includegraphics[width=10cm,clip]{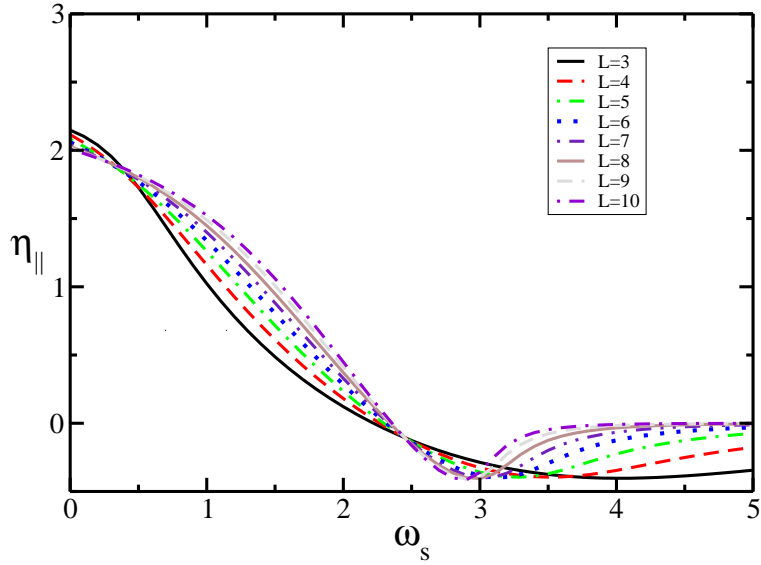}
\end{center}
\end{figure}

\begin{figure}
\caption{Phase diagram calculated by the crossings of $\eta_\parallel$.}\label{eta11pd}
\begin{center}
\includegraphics[width=10cm,clip]{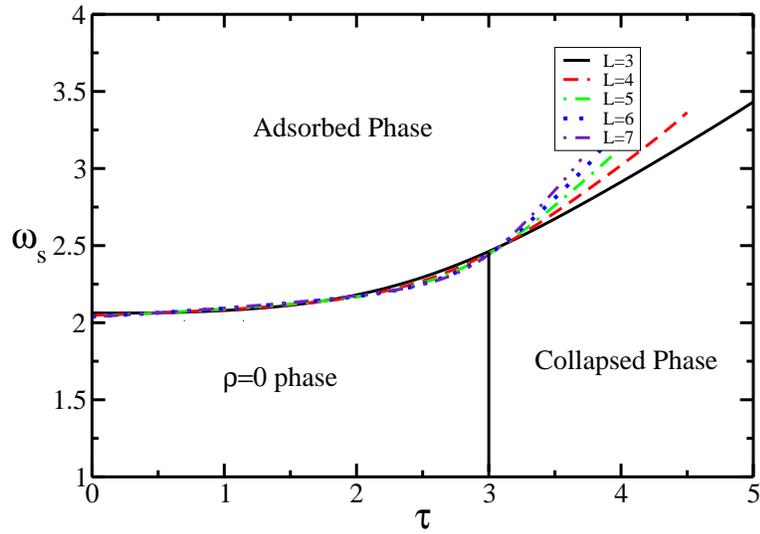}
\end{center}
\end{figure}

In the previous section we calculated our estimate $K^*(\omega_s,\tau)$ by setting $\lambda_1=1$. This gives an overestimate of $K^*$. Calculating $K^*$ using equation~\ref{nrg} gives more asymptotic results, but at the cost of needing two lattice widths, reducing the number of available data points for extrapolation. Additionally, the phenomenological RG method used identifies both the ordinary and the special surface transitions. The phase diagram in the ($K,\omega_s$) plane is shown in figure~\ref{pdkwnrg} for $\tau=3$. It may be seen that there are two points on the phase diagram where the lines seem to cross at (or close to) a point. These two points 
are the fixed points corresponding to the ordinary and the special surface transitions. The location of these points may be estimated using a three width phenomenological  RG scheme. From equations \eref{xi}, \eref{nrg}, \eref{sig-dim} and \eref{eta} it may be seen that we may identify the fixed points with crossings of $\eta_{\parallel}$, and find corresponding estimates for $\eta_{\parallel}$. The plots of $\eta_\parallel$ as a function of $\omega_s$ for $\tau=3$ are shown in \Fref{eta11}, whilst the phase diagram in the $\tau,\omega_s$ plane whith $K=K^*$ is shown in \Fref{eta11pd}. The estimates for the location of the fixed points for the ordinary and  special transitions, along with the corresponding exponent estimates, are shown in \Tref{eta_ord} and \Tref{eta_sp}.

\begin{table}
\caption{Ordinary point location for $\tau=3$ and estimates for $\eta_\parallel^{\rm ord}$. The three point extrapolations are shown in the second half of the table.}\label{eta_ord}
\begin{indented}
\item[]\begin{tabular}{@{}llll}
\br
$L$ &$\omega_s^*$ & $K^*$ & $\eta_\parallel^{\rm ord}$ \\
\mr
3 &0.494271& 0.336986 & 1.734977 \\
4 & 0.369013&  0.334303 & 1.852456\\
5 & 0.313660 &0.333636 &1.892394 \\
6 & 0.289337& 0.333451& 1.907045\\
7 & 0.277118 &0.333387 &1.913419\\
8 & 0.270290 &0.333362 &1.916570\\
9 &  0.266155 &0.333351& 1.918282\\
10 &0.263473 &0.333339 &1.919288\\\mr
BST $\infty$ & 0.256729 & 0.33333 & 1.92133\\\br
\multicolumn{4}{c}{Three point extrapolated results}\\
\mr
3 & 0.225177 & 0.333230 & 1.931118 \\
4 & 0.256335 & 0.333488 & 1.921191 \\
5 & 0.256947 & 0.333401 & 1.921116 \\
6 & 0.256609 & 0.333339 & 1.921221 \\
7 & 0.256325 & 0.333339 & 1.921278 \\
8 & 0.255885 & 0.333409 & 1.921363 \\
\br
\end{tabular}
\end{indented}
\end{table}

\begin{table}
\caption{Special point location for $\tau=3$ and estimates for $\eta_\parallel^{\rm sp}$. The three point extrapolations are shown in the second half of the table.}\label{eta_sp}
\begin{indented}
\item[]\begin{tabular}{@{}llll}
\br
$L$ &$\omega_s^*$ & $K^*$ & $\eta^{\rm sp}_\parallel$ \\
\mr

3 & 2.462062 & 0.332813 & -0.110939 \\
4 & 2.451809 & 0.333074 & -0.105536\\
5 & 2.446565&0.333181 & -0.102198\\
6 & 2.443407& 0.333234&-0.099858 \\
7 & 2.441324 & 0.333264& -0.098106\\
8 & 2.439855 &0.333283& -0.096728 \\
9 & 2.438768 &0.333294 &-0.095606 \\
10 & 2.437932 & 0.333302 & -0.094666\\\mr
BST $\infty$ & 2.43245 & 0.33331 & -0.08449\\
\br
\multicolumn{4}{c}{Three point extrapolated results}\\
\mr
3 &2.434491 &0.333327 &-0.086941 \\
4 & 2.433346 &0.333328 &-0.083971\\
5 &  2.433139& 0.333332 &-0.083208 \\
6 & 2.432852 &0.333340 &-0.082012\\
7 &  2.432683 &0.333317 &-0.081190\\
8 &  2.432473 &0.333339 &-0.079932\\
\br
\end{tabular}
\end{indented}
\end{table}

\begin{table}
\caption{Special point location with $\tau$ unconstrained, using four lattice widths. Estimates for $\eta_\parallel^{\rm sp}$ are also shown. The three point extrapolations are shown in the second half of the table.}\label{eta_sp_uc}
\begin{indented}
\item[]\begin{tabular}{@{}lllll}
\br
$L$ & $\tau^*$ & $\omega_s^*$ & $K^*$ & $\eta^{\rm sp}_\parallel$ \\
\mr
3 &3.132743 & 2.514601  &0.329697&-0.123477\\
4 &3.074773 &2.486484  &0.331186&-0.115417\\
5 &3.048363  &2.472046 &0.331897&-0.110607\\
6 &3.033676 &2.463154  &0.332307 &-0.107244\\
7 &3.024799&2.457278 & 0.332562&-0.104764\\
8& 3.019029&2.453143&  0.332731 &-0.102840\\
9& 3.015082&2.450106& 0.332848  &-0.101298\\
\mr
BST $\infty$ &
2.99967 & 2.440  & 0.333291& -0.0874\\
\br
\multicolumn{4}{c}{Three point extrapolated results}\\
\mr
3 &3.003123& 2.438430 & 0.333265 & -0.091790 \\
4 &2.997885 &2.432266 &0.333423 &-0.084854 \\
5 &  3.000002  &2.433993  &0.333345  &-0.086107 \\
6 & 3.000077 & 2.433675 &0.333338& -0.085025 \\
7 & 3.000388 & 2.433813 &0.333325 & -0.084784 \\
\br
\end{tabular}
\end{indented}
\end{table}

\begin{table}
\caption{Special point location with $\tau=3$ by applying the phenomenological RG method using $x^s_\varepsilon$, leading to estimates for $\nu_s=1/(1-x^s_\epsilon)$.The three point extrapolations are shown in the second half of the table.}\label{Ysurf}
\begin{indented}
\item[]\begin{tabular}{@{}llll}
\br
$L$ &  $\omega_s^*$ & $K^*$ & $\nu_s$ \\
\mr
3  &2.380861 & 0.334825& 1.352725 \\
4 &2.423360 &0.333648  &1.309484 \\
5  &2.439110  &0.333307 &1.291723 \\
6&2.445939 &0.333198  & 1.283136 \\
7&2.449061&0.333167  & 1.278793 \\
8 & 2.450430&0.333166  & 1.276705 \\
9  & 2.450901  &0.333175& 1.275924 \\
10 & 2.450890 &  0.333188  & 1.275943 
\\\mr
BST $\infty$& 2.4512 & 0.3332&1.275\\
\br
\multicolumn{4}{c}{Three point extrapolated results}\\
\mr
3 &  2.456846 &--- & 1.267398 \\
4 & 2.454952  & --- &1.268746 \\
5 &  2.453232    &--- & 1.271502 \\
6 & 2.451999 &--- & 1.273811 \\
7 &  2.451239 &---  &1.275283\\
8& --- &--- & 1.276304\\\br
\end{tabular}
\end{indented}
\end{table}

\begin{table}
\caption{Location of the special point calculated by fixing $K^*$ by phenomenological RG and solving $Y_L(\omega^*_s,\tau)=Y_{L+1}(\omega_s^*,\tau)$ for $\tau=3$. The results are have then been extrapolated using the BST extrapolation scheme.}\label{Ynrg}
\begin{indented}
\item[]\begin{tabular}{@{}llll}
\br
$L$ & $K^*$& $\omega_s^*$ & $Y_L$ \\
\mr
3  &2.675620  &0.326891672428  &0.715973\\
4  &2.622565  &0.329233649135  &0.767296\\
5  & 2.587417 &0.330544853438  &0.809961\\
6  &2.562356  &0.331342978422  &0.845875\\
7  &2.543648  &0.331858273706  &0.876360\\
8  &2.529207  &0.332206642617  &0.902462\\
9  &2.517800  &0.332450587489  &0.924981\\
10 &2.507957  &0.332633569013  &0.945751\\

\mr
BST $\infty$ (3-8) &  0.333377 &  2.427 & 1.19\\
\br
\multicolumn{4}{c}{Three point extrapolated results}\\
\mr
3 & 0.330183 &  2.348631 & --- \\
4 & 0.331123 & 2.369608  & --- \\
5 & 0.333704& 2.388620    & --- \\
6 & 0.333557  & 2.401071 & --- \\
7 &   0.333463 &  2.413110 &--- \\
8 & 0.333647 &2.222366& ---\\\br
\end{tabular}
\end{indented}
\end{table}


The results presented in \Tref{eta_ord} and \Tref{eta_sp} are calculated setting $\tau=3$. The identification $\tau_{\rm coll}$ may be checked by looking for the special transition fixed point in the full phase diagram at the cost of using an extra (fourth) lattice width. Unconstrained results are shown in \Tref{eta_sp_uc}. Extrapolated values of $\tau$ give $\tau_{\rm coll}=3.000\pm0.001$, which is consistent with the identification used elsewhere.  

In the results presented above we have used \eref{nrg}, or equivalently \eref{sig-dim}, to identify the fixed points. However the correlation length defined using the largest and third largest eigenvalues also diverges in the thermodynamic limit, and so we may use a phenomenological RG based on $x_\varepsilon$ using \eref{eng-dim}. This leads to estimates for $\nu_s$, the correlation exponent along the surface as well as alternative estimates for the location of the fixed points. The results are shown in \Tref{Ysurf}.

In \Tref{Ycross} we calculated estimates, ${\cal Y}_L$, of ${\cal Y}_\infty=(1-\phi_s)/\nu$ from \eref{ylog} and \eref{yrg}, calculating $K^*(\omega_s,\tau)$ by setting $\lambda_0=1$. In \Tref{Ynrg} we present the analogous calculation but with $K^*$ calculated using phenomenological RG. 

\subsection{Summary of results}

The transfer matrix calculation gives the exponent $\eta_\parallel$ directly. At the ordinary transition we find a value $\eta_\parallel^{\rm ord}=1.9213\pm 0.0001$. Using the results conjectured in  \cite{F09}, $\nu=12/23$ and $\gamma=22/23$ with \eref{barb} and \eref{gam11eta}, we find 
$\gamma^{\rm ord}_1=0.499$ and $\gamma_{11}^{\rm ord}=-0.481$.
These values agree with the value $\gamma_1^{\rm ord}$ calculated with Monte Carlo in \cite{OP95}, but not with the value of $\gamma_{11}^{\rm ord}$. It should be noted that the values were obtained using $\nu=12/23$ .
At the special transition we find $\eta_\parallel^{\rm sp}=-0.085\pm0.003$, leading to $\gamma_1^{\rm sp}=1.022$ and $\gamma_{11}^{\rm sp}=0.566$. Again, $\gamma_1^{\rm sp}$ is consistent with the value given in \cite{OP95}, but not $\gamma_{11}^{\rm sp}$, and again we used the value $\nu=12/23$. In both cases in \cite{OP95} the values of $\gamma_{11}$ are calculated using the assumed values of $\gamma=1$ and $\nu=1/2$, which we believe not to be the correct values. 

We calculated $\phi_s$ in three ways:
\begin{enumerate}
\item From the result $\nu_s=1.275\pm 0.002$ and using $\nu=12/23$ leads to $\phi_s=\nu/\nu_s=0.414$.
\item Using ${\cal Y}_\infty=(1-\phi_s)/\nu=1.12\pm0.05$ from \Tref{Ycross}, giving $\phi_s=0.405$. 
\item Using ${\cal Y}_\infty=(1-\phi_s)/\nu=1.19\pm0.05$ from \Tref{Ynrg}, giving $\phi_s=0.379$. 
\end{enumerate}
In all three cases the result is close but a little lower than the value $\phi_s=0.440\pm 0.010$ given in \cite{OP95}, but the determination of this exponent is the least accurate of the results presented here. 

\

\section{Discussion}

In this paper we have investigated the surface critical behaviour of the Interacting Self-Avoiding Trail. The results confirm the identification of the collapse transition as occurring at $K_{\rm coll}=1/3$ and $\tau_{\rm coll}=3$ and locates the adsorption transition for $\tau_{\rm coll}=3$ at $\omega_s=2.45\pm0.05$. For the standard Interacting Self-Avoiding Walk model an additional line has been observed in the collapsed phase, separating a region where the collapsed walk is adsorbed to the surface from a region where the walk is desorbed from the surface~\cite{kumar}. This transition corresponds to a wetting of the surface by the perimeter of the collapsed globule, and so does not correspond to a singularity in the bulk free energy. Whilst the method used in this paper does not naturally throw up this transition line, an a-posteriori reexamination of the order parameter defined in \eref{op} for the standard ISAW model 
does show a signature of this transition (see the curves in ref~\cite{veal}).  Such a signature seems to be absent in for this model, see \Fref{opfig} for plots with $\tau=4>\tau_{coll}$. Whilst this should not be taken as a strong indication, it would be of interest to examine whether such a line does exist in this model, or whether the particular type of collapse suppresses this transition.

In conclusion, in the article we use finite-size scaling and transfer matrix methods to investigate the surface critical behaviour of the interacting self-avoiding trail model. Using the conjectured results for $\nu$ and $\gamma$ from reference\cite{F09} ($\nu=12/23$ and $\gamma=22/23$) we find values of $\gamma_1^{\rm ord}$ and $\gamma_1^{\rm sp}$ in agreement with the numerical results of Owczarek and Prellberg\cite{OP95}. We do not, however, find agreement with the conjectured values of $\gamma_{11}^{\rm ord}$ or $\gamma_{11}^{\rm  sp}$. The values of $\nu$ and $\gamma$ needed for this agreement are different from the values $\nu=1/2$ and $\gamma=1$ proposed in \cite{OP95}.

\

\ack

I would like to thank Claire Pinettes for useful discussions and a careful rereading of the manuscript.

\


\end{document}